\documentclass[reprint,author-numerical,nofootinbib]{revtex4-2}
\usepackage{graphicx}
\usepackage{dcolumn}
\usepackage{bm}

\newtheorem{proposition}{Proposition}

\def\dif{{\rm d}}

\newcommand{\be}{\begin{equation}}
\newcommand{\ee}{\end{equation}}

\begin{document}

\preprint{AIP/123-QED}

\title[Thermodynamics of the T-models]
{A thermodynamic approach to the T-models}


\title[Thermodynamics of the T-models]{Thermodynamic approach to the T-models}

\author{Joan Josep Ferrando}
\altaffiliation[Also at ]{Observatori Astron\`omic, Universitat
de Val\`encia,  E-46980 Paterna, Val\`encia, Spain}
\email{joan.ferrando@uv.es.} \affiliation{ Departament d'Astronomia
i Astrof\'{\i}sica, Universitat
de Val\`encia, E-46100 Burjassot, Val\`encia, Spain.}
\author{Salvador Mengual}
\affiliation{ Departament d'Astronomia i Astrof\'{\i}sica,
Universitat de Val\`encia, E-46100 Burjassot, Val\`encia, Spain.}


\begin{abstract}
The perfect fluid solutions admitting a group G$_3$ of isometries acting on orbits S$_2$ whose curvature has a gradient which is tangent to the fluid flow (T-models) are studied from a thermodynamic approach. All the admissible thermodynamic schemes are obtained, and the solutions compatible with the generic ideal gas equation of state are studied in detail. The possible physical interpretation of some previously known T-models is also analyzed.
\end{abstract}

\maketitle

\section{Introduction}
\label{sec-intro}

The spherically symmetric solutions of the Einstein equation have played an essential role in developing the General Relativity theory. But despite the extensive literature on spherically symmetric spacetimes (see, e.g. \cite{Krasinski-Plebanski, kramer, Kim} and references therein), the interest in this topic has not waned nowadays, and several open issues are currently under study. For example, the analyses of the cosmic censorship conjecture have reinforced the study of the properties of the perfect fluid solutions (see \cite{Lapi-Morales, Mosani-J} and references therein). On the other hand, only recently an IDEAL (Intrinsic, Deductive, Explicit and ALgorithmic) characterization of these geometries has been obtained \cite{fs-SSST, fs-SSST-Ricci}, although this kind of approach has been known for longer for some noteworthy solutions \cite{fsS, fsD}.


Static perfect fluid spheres are the most basic and simplest models for studying the stellar structure in both Newtonian and relativistic theories (see, e. g. \cite{Krasinski-Plebanski, Rezzolla}), and the paradigmatic Friedmann-Lema\^itre-Robertson-Walker (FLRW) cosmological models are also spherically symmetric solutions. On the other hand, the most remarkable solution for modeling both gravitational collapse and cosmological inhomogeneities is the Lema\^itre-Tolman model \cite{[{}][{ [English translation: 1997 {\em Gen. Relativ. Gravit.} {\bf 29} 641]}]Lemaitre, [{}][{ [English translation:  1997 {\em Gen. Relativ. Gravit.} {\bf 29} 935]}]Tolman, Krasinski-Plebanski}, which is a spherically symmetric dust solution.

Other non-stationary spherically symmetric perfect fluid spacetimes have been widely considered, and enough families of solutions are known (see \cite{Krasinski-Plebanski, kramer, Krasinski} and references therein). Nevertheless, most solutions have been obtained in the dust case or by prescribing a (non-physical) time-dependence of the pressure, or also by imposing particular barotropic relations. Consequently, further work is required to study the physical meaning of the spherically symmetric perfect fluid solutions, a  study that can also be extended to the plane and hyperbolic symmetries. This task implies analyzing admissible equations of state that fulfill necessary macroscopic constraints for physical reality: energy conditions, compressibility conditions and positivity of some thermodynamic quantities.

We have set ourselves the goal of studying in detail the spacetimes admitting a group G$_3$ of isometries acting on space-like two-dimensional orbits S$_2$ that model the evolution of a thermodynamic perfect fluid in local thermal equilibrium fulfilling the necessary macroscopic constraints for physical reality. Below, we will explain what these macroscopic constraints mean and we present a method to carry them out.


\subsection{Macroscopic necessary conditions for physical reality}
\label{subsec-necessary-conditions}

The evolution of a relativistic perfect fluid is described by an 
energy tensor in the form $T = (\rho+ p) u \otimes u + p \, g$, and fulfilling the conservative condition $\nabla \cdot T=0$. This constraint consists of a differential system of four equations on five {\em hydrodynamic quantities}  ({\em unit velocity} $u$, {\em energy density} $\rho$, and {\em pressure} $p$):
\be \label{ceq}
\hspace{-0mm} {\rm C} :  \quad   \dif p  + u(p) u + (\rho + p) a = 0 \, ,  \quad 
u(\rho) + (\rho+ p) \theta = 0 \, ,
\ee
where $a$ and $\theta$ are, respectively, the acceleration and the
expansion of $u$, and where $u(q)$ denotes the directional
derivative, with respect to $u$, of a quantity $q$, $u(q)
= u^{\alpha} \partial_{\alpha} q$. 

We are interested in perfect energy tensors $T$ that model realistic fluids when the thermodynamic perfect fluid approximation is suitable, that is, when the transport coefficients vanish (or are negligible) \cite{Eckart, Rezzolla}. Next, we summarize the complementary general macroscopic requirements that must be imposed on $T$ to represent the energetic evolution of a physically realistic perfect fluid (see the recent paper \cite{CFS-RSS} for more details).

Pleba\'nski \cite{Plebanski} {\em energy conditions} are necessary algebraic conditions for physical reality and, in the perfect fluid case, they state: 
\begin{equation} \label{e-c}
\hspace{-10mm} {\rm E} :  \qquad \qquad  -\rho < p \leq \rho  \, .
\end{equation}

Furthermore, if we want to describe the (non isoenergetic, $\dot{\rho} \not= 0$) evolution of a thermodynamic perfect fluid in {\em local thermal equilibrium}, the hydrodynamic quantities $\{u, \rho, p\}$ must fulfill the {\em hydrodynamic sonic condition} \cite{Coll-Ferrando-termo, CFS-LTE}: 
\begin{equation} \label{lte-chi}
\hspace{0mm} {\rm S} :  \qquad \quad     \dif \chi \wedge \dif p \wedge \dif \rho = 0 \, , \qquad \chi \equiv \frac{u(p)}{u(\rho)}   \, .
\end{equation}
When this condition holds, the {\em indicatrix of the local thermal equilibrium} $\chi$ is a function of state, $\chi = \chi(\rho,p)$, which physically represents the square of the {\em speed of sound} in the fluid, $\chi (\rho ,p) \equiv  c^2_{s}$. Moreover, a set $\{n, \epsilon, s, \Theta\}$ of {\em thermodynamic quantities} ({\em matter density} $n$, {\em specific internal energy} $\epsilon$, {\em temperature} $\Theta$, and {\em specific entropy} $s$) exists, which is constrained by the common thermodynamic laws \cite{Eckart, Krasinski-Plebanski, Rezzolla}. 
Namely, the conservation of matter:
\begin{equation}  
\nabla \cdot (nu) = u(n) + n \theta = 0 \, ,  \label{c-masa}
\end{equation}
the {\em local thermal equilibrium relation}, which can be written as:
\begin{equation}
\Theta \dif s = \dif h - \frac{1}{n} \dif p \, ,  \qquad h \equiv \frac{\rho+p}{n} \, , \label{re-termo}
\end{equation}
where $h$ is the {\em relativistic specific enthalpy}, and the decomposition defining the specific internal energy:
\begin{equation}
\rho= n(1+\epsilon) \, .  \label{masa-energia} 
\end{equation}
When the conservation equations C and the hydrodynamic sonic condition S hold, we say that $T\equiv \{u, \rho, p\}$ defines the {\em hydrodynamic flow} of a thermodynamic perfect fluid in local thermal equilibrium. Then, the family of {\em thermodynamic schemes} $\{n, \epsilon, s, \Theta\}$ associated with a hydrodynamic flow $T\equiv \{u, \rho, p\}$ is obtained as follows \cite{CFS-LTE}: the specific entropy $s$ and the matter density $n$ are of the form $s= s(\bar{s})$ and $n= \bar{n}R(\bar{s})$, where $s(\bar{s})$ and $R(\bar{s})$ are arbitrary real functions of a particular solution $\bar{s}=\bar{s}(\rho, p)$ to the equation $u(s)=0$, and $\bar{n}=\bar{n}(\rho,p)$ is a particular solution to the equation (\ref{c-masa}). Moreover, $\Theta$ and $\epsilon$ are determined, respectively, by (\ref{re-termo}) and (\ref{masa-energia}).

A basic physical requirement imposed on the thermodynamic schemes is the positivity of the matter density, of the temperature and of the specific internal energy,
\begin{equation} \label{P}
\hspace{-5mm} {\rm P} :  \qquad \qquad     \Theta > 0 \, , \qquad  \quad   \rho > n > 0   \, .
\end{equation}

Finally, in order to obtain a coherent theory of shock waves for the fundamental system of perfect fluid hydrodynamics \{(\ref{ceq}),(\ref{c-masa}),(\ref{re-termo}),(\ref{masa-energia})\} one must impose the relativistic compressibility conditions  \cite{Israel, Lichnero-1, Anile, Lichnero-2}. They impose the inequalities ${\rm H}_1: \,   (\tau'_p)_s < 0, \ (\tau''_p)_s > 0$, and the inequality ${\rm H}_2: \, (\tau'_s)_p > 0$, where the function of state $\tau = \tau(p, s)$ is the {\em dynamic volume}, $\tau = \hat{h}/n$, $\hat{h} = h/c^2$ being the dimensionless enthalpy index. In \cite{CFS-CC} we have shown that the compressibility conditions H$_1$ only restrict the hydrodynamic quantities, and that they can be stated in terms of the function of state $c_s^2 = \chi(\rho,p)$: 
\begin{equation}
\hspace{0mm} {\rm H}_1 :  \quad 0 < \chi < 1 , \ \  \  (\rho+p)(\chi \chi_{p}' + \chi_{\rho}') + 2 \chi(1-\chi) > 0    .       \label{cc-1-chi}
\end{equation}
However, compressibility condition H$_2$ imposes constraints on the thermodynamic scheme and it can be stated as \cite{CFS-CC}:
\be \label{H2-Theta}
\hspace{-5mm} {\rm H}_2 : \qquad \qquad  2 n \Theta > \frac{1}{s_{\rho}'}    \, .
\ee
%


\subsection{Procedure to determine physically admissible perfect fluid solutions }
\label{subsec-physical-solutions}

Note that in the general necessary macroscopic constraints C, E, S, P, H$_1$ and H$_2$ specified above, we must distinguish two types of conditions according to their nature:
\begin{itemize}
\item[a)] 
{\em Hydrodynamic constraints}: the conservation equation C, the energy conditions E, the hydrodynamic sonic condition S, and the compressibility conditions H$_1$ exclusively involve the hydrodynamic quantities $\{u, \rho, p\}$. They fully determine the hydrodynamic flow of the thermodynamic fluid in local thermal equilibrium and, consequently, restrict the admissible gravitational field as a consequence of the Einstein equations.
\item[b)]
{\em Thermodynamic constraints}: the positivity conditions P and the compressibility condition H$_2$ restrict the thermodynamic schemes $\{n, \epsilon, s, \Theta\}$ associated with a hy\-dro\-dynamic flow $\{u, \rho, p\}$. Consequently, they do not restrict the gravitational field and the admissible thermodynamics offer different physical interpretations for a given hydrodynamic perfect fluid flow. 
\end{itemize}

In order to implement the above macroscopic constraints in looking for physically admissible new perfect fluid solutions and in analyzing the previously known ones, we have proposed in \cite{CFS-RSS} a general procedure in five steps:
\begin{description}
\item[{\bf Step 1}]
Determine the subfamily of the thermodynamic solutions by imposing the hydrodynamic sonic condition S on the solutions to the conservative equations C. 
\item[{\bf Step 2}]
Obtain, for this subfamily, the coordinate dependence of the hydrodynamic quantities $u$, $\rho$, $p$, and  the indicatrix function $c_s^2 = \chi(\rho,p)$.
\item[{\bf Step 3}]
Analyze, for these thermodynamic solutions, the hydrodynamic constraints for physical reality, namely, the energy conditions E and the compressibility conditions H$_1$.
\item[{\bf Step 4}]
Obtain the thermodynamic schemes $\{n, \epsilon, s, \Theta\}$ associated with these solutions. \item[{\bf Step 5}]
Analyze, for the thermodynamic schemes $\{n, \epsilon, s, \Theta\}$ already obtained, the general thermodynamic constraints for physical reality, namely, the positivity conditions P and the compressibility condition H$_2$.
\end{description}
This procedure or an adapted version thereof has been used elsewhere in studying the ideal gas Stephani universes \cite{CFS-CC, CF-Stephani}, the classical ideal gas solutions \cite{CFS-CIG} and the singular and the regular models of the thermodynamic class II Szekeres-Szafron solutions \cite{CFS-RSS, CFS-PSS}. Our further study of the physical reality of the spacetimes admitting a group G$_3$ of isometries on orbits S$_2$ will be also based on this approach.


\subsection{About this paper}
\label{subsec-thispaper}
%


In comoving-synchronous coordinates, the metric of a perfect fluid solution admitting a three-dimensional group G$_3$ of isometries acting on spacelike two-dimensional orbits S$_2$ has the form \cite{kramer}:
\begin{eqnarray} \label{metric-ss-1}
ds^2= -e^{2\nu}dt^2 + e^{2\lambda} dr^2 + Y^2 C^2 (dx^2+dy^2),  \ \\[3mm]
%
%
\nu=\nu(r, t), \qquad \lambda=\lambda(r, t), \qquad Y=Y(r, t),  \ \label{metric-ss-2} \\[1mm]
%
C=C(x, y)\equiv\left[1+\frac{k}{4}(x^2+y^2)\right]^{-1} \!\!\!\!, \quad k=0, \pm1 , \ \label{metric-ss-3}
\end{eqnarray}
where the value of $k$ distinguishes the plane, spherical and hyperbolic symmetries.

The $r$-dependence of the functions $\nu$ and $Y$ plays an important role in the analysis of the Einstein equations for a perfect energy tensor source. Thus, usually one considers separately the cases $\nu = \nu(t)$ (geodesic motion) or/and $Y=Y(t)$ (T-models) \cite{Krasinski-Plebanski, kramer}. On the other hand, Ruban \cite{Ruban} showed that the spherically perfect fluid T-models have geodesic motion (see also \cite{Krasinski}), a result that can be extended to the plane and hyperbolic symmetries (see, for example, \cite{Krasinski-Plebanski}). Hereon in, {\em T-models} refer to the perfect fluid solutions whose metric has the form (\ref{metric-ss-1}-\ref{metric-ss-3}) with $\nu=\nu(t)$ and $Y=Y(t)$. 

The spherical dust T-model was published in a pioneer paper by Datt \cite{[{}][{ [English translation: 1999 {\em Gen. Relativ. Gravit.} {\bf 31} 1619]}]Datt} and rediscovered later by Ruban \cite{[{}][{ [English translation: 1968 {\it Sov. Phys. JETP Lett.} {\bf 8} 414] [Reprinted: 2001 {\em Gen. Relativ. Gravit.} {\bf 33} 363]}]Ruban-68, [{}][{ [English translation: 1969 {\it Sov. Phys. JETP} {\bf 29} 1027] [Reprinted: 2001 {\em Gen. Relativ. Gravit.} {\bf 33} 375]}]Ruban}, and the general perfect fluid solution with a non-constant pressure was considered by Korkina and Martinenko \cite{Korkina}. An exhaustive list of the particular solutions presented by several authors can be found in \cite{Krasinski}, but the physical meaning of any of these solutions is doubtful. 

The geometric and physical properties of the dust T-models were analyzed by Ruban \cite{Ruban-68} (see also \cite{Krasinski-Plebanski}). The metric is invariant under the group of rotations but, since $Y$ depends only of $t$, the space-like 3-spaces $t= constant$ do not contain their center of symmetry. The geometry of these 3-spaces is that of a three-dimensional cylinder, that is, the direct product of a 2-sphere and an open straight line. A similar situation occurs in the flat or the hyperbolic symmetries by changing the 2-sphere by a plane or a hyperboloid. Moreover, extensive work (see \cite{Bonnor-ST, G-hellaby, Krasinski, Krasinski-Plebanski} and references therein) has been devoted to extended this study to the Szekeres solutions of class II \cite{Szekeres}, which are their generalizations without symmetries. The solutions with non-constant pressure basically keep the geometric properties of the dust solutions \cite{[{}][{ [English translation: 1983 {\it Sov. Phys. JETP} {\bf 58} 463]}]Ruban-83}, but the physical meaning of these T-models is still an open problem.

Here, we present a thermodynamic approach to the T-models. It is worth remarking that these metrics define a subfamily of the class II Szekeres-Szafron solutions \cite{Krasinski-Plebanski, Krasinski, Szekeres, Szafron, FS-SS}, the only one left to study from a thermodynamical approach. The other thermodynamic Szekeres-Szafron solutions of class II, the singular and the regular models, have recently been studied elsewhere \cite{CFS-PSS, CFS-RSS}.

In Section \ref{sec-Tmodels}, we revisit the perfect fluid field equations for the T-models and we show that they can be formulated as ordinary differential equations which are linear for a suitable choice of the unknown metric functions.

Section \ref{sec-thermo-Tmodels} is devoted to analyzing the general thermodynamic properties of the T-models by obtaining the hydrodynamic quantities and the hydrodynamic equation of state $c_s^2=\chi(\rho, p)$. All the possible thermodynamic interpretations of each solution of the field equations are also presented by obtaining all the compatible thermodynamic schemes.

In Section \ref{sec-chi-pi} we study the T-models that are compatible with the equation of state of a generic ideal gas, $p= \tilde{k} n \Theta$, to which we apply our procedure to analyze the physical reality of the solutions. We show that some of the solutions demonstrating good physical behavior belong to the Szekeres-Szafron ideal singular models considered in \cite{CFS-PSS}. The physical behavior of some thermodynamic schemes is analyzed.

It is known \cite{Krasinski} that the spatially homogeneous limit of the T-models, $\lambda= \lambda(t)$, are the Kompanneets-Chernov-Kantowski-Sachs (KCKS) metrics \cite{Kompa, K-S}. These models were considered by Kompanneets and Chernov \cite{[{}][{ [English translation: 1965 {\em Sov. Phys. JETP} {\bf 20} 1303]}]Kompa} and were studied by Kantowski and Sachs \cite{K-S} for a dust source. One particular solution in this family was obtained and analyzed by McVittie and Wiltshire \cite{McVittie}, and later generalized by Herlt \cite{Herlt} for the inhomogeneous case. In Section \ref{sec-McVittie} we generalize this McVittie-Wiltshire-Herlt solution to any curvature and we discuss the unclear physical interpretation of this solution.

Finally, in Section \ref{sec-discussion} we point out our results, we remark on the constraints in looking for solutions that model a classical ideal gas, and we comment on our ongoing work.


\section{T-models: field equations for the metric functions}
\label{sec-Tmodels}

If we make $e^{\lambda} = \omega(t,r)>0$, $e^{-2 \nu} = v(t)>0$ and $Y^2 = \varphi(t)>0$ in the metric line element (\ref{metric-ss-1}) it follows that the metric tensor of a T-model can be written as:
\begin{equation} \label{metric-T-1}
ds^2= -\frac{1}{v(t)}dt^2 + \omega^2(t,r) dr^2 + \varphi(t) C^2 (dx^2+dy^2),
\end{equation}
where $C$ is given in (\ref{metric-ss-3}). Moreover, from the general expressions for the field equations for the metric (\ref{metric-ss-1}) (see, for example, \cite{Krasinski-Plebanski, kramer}), it follows that (\ref{metric-T-1}) is a perfect fluid solution if, and only if, the metric functions $v(t)$, $\omega(t,r)$ and $\varphi(t)$ meet the differential equation:
\be \label{eq-T-1}
2 v \varphi \, \ddot{\omega} + (\dot{v} \varphi + v \dot{\varphi})\, \dot{\omega} - (v \ddot{\varphi} + \frac12 \dot{v} \dot{\varphi} + 2k)  \,\omega = 0 \, ,
\ee
where a dot denotes derivative with respect to the time coordinate $t$.

The unit velocity of the fluid $u = \sqrt{v} \, \partial_t$ is geodesic and its expansion is:
\begin{equation} \label{expansion-T-1}
\theta = \sqrt{v} \left(\frac{\dot{\varphi}}{\varphi} + \frac{ \dot{\omega}}{\omega}\right)  = \sqrt{v} \, \partial_t [\ln(\varphi \omega)] \,  .
\end{equation}
And the {\em pressure} $p$ and the {\em energy density} $\rho$ are then given by: 
\begin{eqnarray} \label{pressure-T-1}
p =  v \left[\frac14 \frac{\dot{\varphi}^2}{\varphi^2} - \frac{\ddot{\varphi}}{\varphi} - \frac12 \frac{\dot{\varphi}}{\varphi}  \frac{ \dot{v}}{v}\right] - \frac{k}{\varphi}   \,  ,  \\[1mm]
\rho =  v \left[\frac14 \frac{\dot{\varphi}^2}{\varphi^2} +  \frac{\dot{\varphi}}{\varphi}  \frac{\dot{\omega}}{\omega}\right] + \frac{k}{\varphi}   \,  .
\label{density-T-1}
\end{eqnarray}

The spatially homogeneous limit of the T-models are the KCKS metrics \cite{Kompa, K-S}, which admit a group G$_4$ of isometries acting on orbits S$_3$. They can also be characterized by one of the following three equivalent conditions: (i) the metric function $\omega(t,r)$ factorizes, and then one can take the coordinate $r$ so that $\omega= \omega(t)$, (ii) the energy density is homogeneous, $\rho = \rho(t)$, and (iii)  the fluid expansion is homogeneous, $\theta = \theta(t)$.

Note that (\ref{eq-T-1}) is a second order linear differential equation for the function $\omega(t,r)$ when $v(t)$ and $\varphi(t)$ are given. Consequently, its general solution is of the form $\omega(t,r) = \omega_1(t) Q_1(r) + \omega_2(t) Q_2(r)$. Moreover, we can change the coordinate $r$ so that:
\be \label{w-w1-w2}
\omega(t,r) =  \omega_1(t) + \omega_2(t) \, Q(r) \, ,
\ee
where $Q(r)$ is an arbitrary real function, and with $\omega_i(t)$ being two particular solutions to the equation (\ref{eq-T-1}). 

We have the freedom to choose the coordinate $t$ without changing the spacetime metric. Therefore, we can impose a condition on the time-dependent functions $v$, $\varphi$ and $\omega_i$ that fixes this election. Consequently, the space of solutions depends on two arbitrary real functions, one depending on $r$, $Q(r)$, and the other one depending on time.

For example, if we take $v(t)=1$, then the coordinate $t$ is the proper time of the co-moving observer. In this case, for every choice of the function $\varphi(t)$, equation (\ref{eq-T-1}) determines two particular solutions $\omega_i(t)$. Thus, the space of solutions is controlled by the functions $\{\varphi(t), Q(r)\}$.

On the other hand, it is quite common in the literature (see, for example, \cite{kramer, Krasinski}) to consider $t= Y = \sqrt{\varphi}$. Then, if we give $v(t)$, the functions $\omega_i(t)$ are determined by equation (\ref{eq-T-1}), and thus the space of solutions is controlled by the functions $\{v(t), Q(r)\}$. But we can also give as input one of the functions $\omega_i$, and then equation (\ref{eq-T-1}) becomes a first order linear differential equation for the function $v(t)$; once this equation is solved, we can proceed to determine the other $\omega_i$ by once again using (\ref{eq-T-1}) with the $v(t)$ previously obtained. This procedure has been used, for example, by Herlt \cite{Herlt}. 

It is worth remarking that our choice of the metric function $\varphi = Y^2$ as an unknown of the field equations, leads us to equation (\ref{eq-T-1}), which is also a linear equation for $\varphi$. Then, this equation is linear for the three involved metric functions, a fact that may certainly be of interest to a further search for new solutions. 

The solutions known so far have been obtained by prescribing the free metric functions in a way that allows analytical integration of field equations \cite{Krasinski-Plebanski, kramer, Herlt}, but without any evident physical meaning. Therefore, it seems of interest to study the thermodynamic interpretation of the known solutions, as well as to obtain new solutions that meet some previously prescribed physical properties.


\section{Thermodynamics of the T-models}
\label{sec-thermo-Tmodels}

Now we analyze when the T-models (\ref{metric-T-1}-\ref{eq-T-1}) represent the evolution in local thermal equilibrium of a fluid that meets the suitable macroscopic physical constraints stated in subsection \ref{subsec-necessary-conditions}. Note that the existing symmetries imply that all the scalar invariants, and in particular the energy density $\rho$, the pressure $p$ and the indicatrix function $\chi$, depend on two functions at most. Then, the sonic condition S given in (\ref{lte-chi}) identically holds and, consequently, step 1 in the procedure presented in subsection \ref{subsec-physical-solutions} is achieved for the full set of T-models. Thus, we proceed to analyze step 2.


\subsection{Metric and hydrodynamic quantities: unit velocity, energy density and pressure} 
\label{subsec-metric-Tmodels}

In the previous section we have already given the metric and the hydrodynamic quantities of the T-models. Now, for the sake of simplicity and in order to facilitate the calculation in studying the thermodynamic properties, we choose $v=1$. This means that the time coordinate $t$ is the proper time of the Lagrangian observer associated with the fluid. Then, it follows that the metric tensor of the perfect fluid T-models can be written as:
\begin{equation} \label{metric-T-2}
ds^2= -dt^2 + [\omega_1(t) + \omega_2(t) \, Q(r)] ^2 dr^2 + \varphi(t) C^2 (dx^2+dy^2),
\end{equation}
where $C$ is given in (\ref{metric-ss-3}), and $\omega_i(t)$ are two particular solutions of the second order differential equation:
\be \label{eq-T-2}
2 \varphi \, \ddot{\omega} +  \dot{\varphi}\, \dot{\omega} - (\ddot{\varphi} + 2k)  \,\omega = 0 \, .
\ee

The unit velocity of the fluid $u =  \partial_t$ is geodesic and its expansion is:
\begin{equation} \label{expansion-T-2}
\theta = \frac{\dot{\varphi}}{\varphi} + \frac{ \dot{\omega}}{\omega}  = \partial_t (\ln[\varphi (\omega_1 + \omega_2 Q)])\,  .
\end{equation}
And the {\em pressure} $p$ and the {\em energy density} $\rho$ are then given by: 
\begin{eqnarray} \label{pressure-T-2}
p =  \frac14 \frac{\dot{\varphi}^2}{\varphi^2} - \frac{\ddot{\varphi}}{\varphi} - \frac{k}{\varphi}     \,  ,  \\[1mm]
\rho =  \frac14 \frac{\dot{\varphi}^2}{\varphi^2} +  \frac{\dot{\varphi}}{\varphi} \,  \frac{\dot{\omega}_1 + \dot{\omega}_2 Q} { \omega_1 + \omega_2 Q} + \frac{k}{\varphi}   \,  .
\label{density-T-2}
\end{eqnarray}

Note that, with our choice $v(t)=1$, the space of solutions of the T-models depends on the real functions $\{\varphi(t), Q(r)\}$. Moreover, the barotropic limit (KCKS metrics) is achieved when $Q(r)=constant$. These barotropic models may represent an isentropic evolution of a thermodynamic fluid \cite{CFS-LTE} (see section \ref{sec-discussion}). On the other hand, solutions in local thermal equilibrium with constant pressure lead, necessarily, to an isobaroenergetic evolution, $\dot{\rho} = \dot{p} =0$ \cite{CFS-LTE}; then, the fluid expansion vanishes as a consequence of (\ref{ceq}), and (\ref{expansion-T-2}) implies that $\omega$ factorizes and the metric is a degenerate KCKS model. Moreover, if $\dot{\varphi} =0$ then (\ref{pressure-T-2}) and (\ref{density-T-2}) imply that $\rho+p=0$ and the energy conditions (\ref{e-c}) do not hold. From now on, in this paper, we will consider the T-models (\ref{metric-T-2}-\ref{eq-T-2}) with $Q'(r) \not=0$, $\dot{p}(t) \not=0$ and $\dot{\varphi} \not=0$.


\subsection{The indicatrix function: speed of sound}
\label{subsec-chi-Tmodels}

In order to simplify calculations we define the following functions:
\begin{equation} \label{abcd}
\begin{array}{l}
\displaystyle \sigma(t) \equiv \frac{\ddot{\varphi}}{\varphi} \, ,  \quad  \beta(t) \equiv \frac{\dot{\varphi}^2}{\varphi^2} \, ,  \quad \xi(t) \equiv \frac{k}{\varphi} \,  , \\[4mm]   \displaystyle \Omega(t,r) \equiv \frac{\dot{\omega}_1 + \dot{\omega}_2 Q} { \omega_1 + \omega_2 Q}\ \, .
\end{array}
\end{equation}
Then, the pressure, the energy density and the expansion take the form:
\begin{equation} \label{p-density-expansion}
\hspace{0cm} p = \frac14 \beta - \sigma -  \xi   , \quad  \rho =  \frac14 \beta + \xi +  \sqrt{\beta} \,  \Omega  , 
\quad \theta = \sqrt{\beta} + \Omega  .
\end{equation}

Now we can calculate the square of the speed of sound in terms of the hydrodynamic quantities $\rho$ and $p$ by using the expression (\ref{lte-chi}) of the indicatrix function. Note that now, with the choice $v=1$, we have $u(q) = \dot{q}$ for any scalar quantity $q$. From the equation (\ref{ceq}) and the expression of the expansion (\ref{p-density-expansion}), we obtain:
\begin{equation} \label{rho-punt}
u(\rho) = \dot{\rho} = -\frac{1}{\sqrt{\beta}} [\rho^2 + (p+q) \rho + pq]  , \quad q \equiv \frac34 \beta - \xi  .
\end{equation}
Consequently, we have the following result:
\begin{proposition} \label{prop-chi-Tmodels}
For the T-models {\em (\ref{metric-T-2}-\ref{eq-T-2})}, the square of the speed of sound takes the expression:
\be  \label{chi-Tmodels}
c_s^2  = \frac{u(p)}{u(\rho)} = \chi(\rho,p)  \equiv \frac{1}{{\cal A}(p) \rho^2 + {\cal B}(p) \rho + {\cal C}(p)} \, ,
\ee
where ${\cal A}$, ${\cal B}$ and ${\cal C}$ are the functions of $t$ (and then of $p$) given by:
\begin{equation}  
\hspace{0mm} {\cal A}(p) \equiv  -\frac{1}{\sqrt{\beta} \dot{p}}  , \quad {\cal B}(p) \equiv  {\cal A} \, (p + q)  ,  \quad {\cal C}(p) \equiv  {\cal A} \, p \,q  \label{ABCcal}   .   
\end{equation}
\end{proposition}
It is worth remarking that the expression (\ref{chi-Tmodels}) for the square of the speed of sound is similar to that obtained for the singular and regular models of the thermodynamic Szekeres-Szafron solutions of class II \cite{CFS-PSS, CFS-RSS}. This fact was to be expected since the T-models define the subfamily of this class that we had left to study from a thermodynamic approach.

The equation of state $c_s^2 = \chi(\rho,p)$ given in (\ref{chi-Tmodels}) collects all the thermodynamic information that can be expressed using exclusively hydrodynamic quantities. Note that the dependence on the variable $\rho$ is explicit, but the dependence on $p$ is implicit through the functions ${\cal A}$, ${\cal B}$ and ${\cal C}$ given in (\ref{ABCcal}). These functions only depend on $\varphi(t)$ and its derivatives. Thus the explicit form of $\chi(\rho,p)$ may be obtained when a specific $\varphi(t)$ is given (see the following sections).

Once steps 1 and 2 of the procedure proposed in subsection \ref{subsec-physical-solutions} have been achieved, we could formally impose the restrictions required in step 3 (energy and compressibility conditions H$_1$). Nevertheless, we delay this study for subclasses of solutions that fulfill complementary physical requirements, and once we have obtained the explicit form of $\chi(\rho,p)$. Now we analyze step 4 for the whole set of T-models.


\subsection{Thermodynamic scheme: entropy, matter density and temperature} 
\label{subsec-scheme-Tmodels}

In this subsection we solve the inverse problem \cite{CFS-LTE} for the T-models (\ref{metric-T-2}-\ref{eq-T-2}) by obtaining the full set of thermodynamic quantities: specific entropy $s$, matter density $n$ and temperature $\Theta$. The metric function $Q(r)$ plays an important role in this thermodynamic scheme. From the expressions given in (\ref{abcd}) and (\ref{p-density-expansion}) we obtain:
\begin{equation} \label{Q(prho)}
Q =- \frac{(\rho-p) \varphi \, \omega_1 - {(\dot{\varphi} \omega_1)}^{\cdot}-2k \omega_1}{(\rho-p) \varphi \, \omega_2 - {(\dot{\varphi} \omega_2)}^{\cdot}-2k \omega_2}  \equiv Q(\rho,p) \, .
\end{equation}
Note that $Q=Q(\rho,p)$ is a function of state whose dependence on $\rho$ is explicit, while its dependence on $p$ is partially implicit through the functions of time $\omega_i(t)$ and $\varphi(t)$. We have that $\dot{Q}=0$ and, consequently, $Q$ is a particular solution of $u(s) = 0$. 

On the other hand, from the expression (\ref{expansion-T-2}) of the expansion it follows that $\bar{n} = [\varphi(\omega_1 + \omega_2 Q)]^{-1}$ is a particular solution of the matter conservation equation (\ref{c-masa}). Then, taking into account the thermodynamic view presented in subsection \ref{subsec-physical-solutions}, we obtain the following:
\begin{proposition} \label{prop-s-n-Tmodels}
The thermodynamic schemes associated with the T-models {\em (\ref{metric-T-2}-\ref{eq-T-2})} are determined by a specific entropy $s$ and a matter density $n$ of the form:

\begin{equation}  \label{s-n-Tmodels}
s(\rho, p) = s(Q)    ; \quad  n(\rho,p) = \frac{1}{\varphi(\omega_1 + \omega_2 Q)N(Q)}   ,
\end{equation}
where $s(Q)$ and $N(Q)$ are two arbitrary real functions.
\end{proposition}

The temperature of the thermodynamic scheme defined by each pair $\{s, n\}$ given in proposition above can be obtained from the thermodynamic relation (\ref{re-termo}). Expressions (\ref{p-density-expansion}) and (\ref{s-n-Tmodels}) imply that the specific enthalpy is:
\begin{eqnarray}  \label{h-Tmodels}
h = \frac{\rho+p}{n} =  N(Q)[\lambda_1(t) + Q \lambda_2(t)]  ,  \qquad \quad  \\  
\lambda_i(t) \equiv \dot{\varphi} \,\dot{\omega}_i + \left[\frac{\dot{\varphi}^2}{2 \varphi} - \ddot{\varphi}\right] \omega_i = 2 Y(\dot{Y} \dot{\omega}_i - \ddot{Y} \omega_i)  , \ \  \label{lambda_i}
\end{eqnarray}
where $Y = \sqrt{\varphi}$. Then, from (\ref{re-termo}) we have $\Theta = \left(\frac{\partial h}{\partial s}\right)_p = \frac{1}{s'(Q)}  
\left(\frac{\partial h}{\partial Q}\right)_t$ and, taking into account (\ref{h-Tmodels}), we obtain the following: 
\begin{proposition} \label{prop-T-Tmodels}
For the T-models {\em (\ref{metric-T-2}-\ref{eq-T-2})}, the temperature $\Theta$ of the thermodynamic schemes given in proposition {\em \ref{prop-s-n-Tmodels}} takes the expression:
\begin{equation}  \label{T-Tmodels}
\Theta = \ell(Q) \lambda_1(t) + m(Q) \lambda_2(t) \equiv \Theta(\rho,p)  \, , 
\end{equation}
where $\lambda_i(t)$ is given in {\em (\ref{lambda_i})} and 
\begin{equation}
 \ell(Q) \equiv \frac{N'(Q)}{s'(Q)}  , \quad  m(Q) \equiv \frac{1}{s'(Q)}[Q N'(Q) + N(Q)]      .  \label{ell-m}
\end{equation}
\end{proposition}

The last step of the procedure presented in subsection \ref{subsec-physical-solutions} consists in the study of the compatibility of the thermodynamic schemes above considered with the positivity conditions P and the compressibility condition H$_2$. This analysis will be efficient when we consider a specific solution and we may obtain all the thermodynamic quantities in terms of the hydrodynamic ones $\rho$ and $p$ (see the following sections).


\section{T-models compatible with the equation of state of a generic ideal gas}
\label{sec-chi-pi}


Now we will analyze when the T-models and the associated thermodynamic schemes considered above are compatible with the equation of state of a {\em generic ideal gas}, namely:
\begin{equation}
p = \tilde{k} n \Theta  \, , \qquad \quad    \tilde{k} \equiv {k_B \over m} \,  .  \label{gas-ideal}
\end{equation}
In \cite{CFS-LTE} we have shown that equation (\ref{gas-ideal}) restricts the functional dependence of the indicatrix function $c_s^2   = \chi(\rho,p)$. More precisely: a perfect energy tensor $T=\{u,\rho,p\}$ represents the evolution of a generic ideal gas in local thermal equilibrium if, and only if, it fulfills the {\em ideal gas sonic condition}:
\be \label{chi-gas-ideal}
\hspace{-2mm} {\rm S^{\rm G}} :  \quad \quad     \chi = \chi(\pi) \not= \pi  , \quad \chi \equiv \frac{u(p)}{u(\rho)}  , \quad \pi \equiv \frac{p}{\rho}  .
\ee

On the other hand, in \cite{CFS-CC} we have proved that, for an indicatrix function of the form (\ref{chi-gas-ideal}), $\chi = \chi(\pi)$, the compressibility conditions H$_1$ given in (\ref{cc-1-chi}) become:
\begin{equation}
\hspace{0mm} {\rm H}_1^{\rm G} :  \ \        0 < \chi < 1  , \ \   \zeta \equiv (1+\pi)(\chi-\pi) \chi'  + 2 \chi(1-\chi) > 0    .       \label{cc-ideal}
\end{equation}

Moreover, the equation of state (\ref{gas-ideal}) and the positivity conditions P given in (\ref{P}) imply a non-negative thermodynamic pressure, $p > 0$. Consequently, the energy conditions E given in (\ref{e-c}) become (here we shall consider non-shift perfect fluids, $\rho \not=p$):
\begin{equation} \label{e-c-G}
\hspace{-5mm} {\rm E}^{\rm G} : \qquad \quad    \rho > 0 \, , \qquad  0 < \pi < 1 \, , \quad  \pi = \frac{p}{\rho} \, .
\end{equation}

Note that, in order to study the solutions with the hydrodynamic behavior of a generic ideal gas, we can 
fairly modify the procedure exposed in subsection \ref{subsec-physical-solutions} by changing the sonic condition S, the energy condition E and the compressibility conditions H$_1$ by the corresponding S$^{\rm G}$, E$^{\rm G}$ and H$_1^{\rm G}$.    


\subsection{Study of the ideal sonic condition {\rm S}$^{\rm G}$}
\label{subsec-ideal-equations}

We must study the modified step 1 by analyzing which T-models meet the ideal sonic condition S$^{\rm G}$ given in (\ref{chi-gas-ideal}).  From the expression of the indicatrix function (\ref{chi-Tmodels}) it follows that (\ref{chi-gas-ideal}) is equivalent to:
\be \label{ABCci}
{\cal A} p^2 = c_1 \, , \quad {\cal B} p = c_2 \, , \quad {\cal C} = c_3 \, , \quad c_i = constant \, .
\ee
Then, expressions (\ref{ABCcal}) lead to:
\begin{equation} 
c_1 = -  \frac{p^2}{\sqrt{\beta} \dot{p}} \not=0 \, , \quad c_2 = \frac{c_1}{p}(p+q)  \, ,  \quad c_3 = \frac{c_1}{p}q  \, ,  \label{ideal-eq}     
\end{equation}
or, equivalently:
\be
c_2 = c_1 + c_3 \, , \qquad c_1\, q = c_3\, p  \, , \qquad \dot{p} = - \frac{p^2}{c_1 \sqrt{\beta}}. \label{ideal-eq-b}
\ee

If $c_3=0$, (\ref{ideal-eq-b}) implies $q=0$ and then $\beta = \frac34 \xi$, that is, function $\varphi(t)$ fulfills an equation of the form $\dot{\varphi}^2 = c_0^2 \varphi$. Otherwise, if $c_3 \not=0$, from the first equation in (\ref{p-density-expansion}) and the second equation in (\ref{ideal-eq-b}), and taking into account the definitions of $\sigma$, $\beta$, $\xi$ given in (\ref{abcd}) and of $q$ given in (\ref{rho-punt}), we obtain:
\be \label{betapdot}
\begin{array}{c}
\dot{\beta} = 2 \sqrt{\beta}\Big[\Big(\frac{c_3}{c_1}-1\Big)p - \frac32 \beta \Big] , \\[3mm]
 \dot{p} = \frac12 \sqrt{\beta} \Big[\Big(1-3 \frac{c_1}{c_3}\Big)p - 3 \frac{c_1}{c_3} \beta \Big] .
 \end{array}
\ee
Then, the two expressions (\ref{ideal-eq-b}) and (\ref{betapdot}) for $\dot{p}$ lead to: 
\be \label{beta-p-2}
2 p^2 + c_1 \Big(1-3 \frac{c_1}{c_3}\Big) p \beta - 3 \frac{c_1^2}{c_3} \beta^2 = 0 \, .
\ee
From the second expression in (\ref{ideal-eq-b}) and the definition (\ref{rho-punt}) of $q$ we obtain $p = \frac{c_1}{c_3}(\frac34 \beta -\xi)$. Then, we can substitute $p$ and equation (\ref{beta-p-2}) becomes:
\be \label{beta-xi}
2 \xi^2 + (3 c_1 - c_3 - 3)  \beta \xi  + \frac98 (1 - 2c_1 - 2c_3) \beta^2 = 0 \, .
\ee
This equation is a necessary constraint for the compatibility of the ideal sonic condition. Now we consider two cases. If $k=0$, then $\xi=0$, and (\ref{beta-xi}) states $2(c_1 + c_3) =1$. Otherwise, if $k \not=0$, for a given $\beta$, (\ref{beta-xi}) is a second degree algebraic equation for $\xi$ that must admit solution. Then, this solution is of the form $\xi = b_0 \beta$, $b_0 = constant\not=0$. Consequently, the ideal sonic condition admits a solution if one of the two following conditions holds:
\begin{itemize}
\item[i)]
$k=0$, and $c_2 = c_1 + c_3 = \frac12$.
\item[ii)]
$k \not= 0$, and $\dot{\varphi}^2 = c_0^2 \, \varphi$, $c_0 = constant$. 
\end{itemize}
Case (ii) leads to negative pressures and is not compatible with the generic ideal gas equation of state (\ref{gas-ideal}). It will be analyzed in section \ref{sec-McVittie}. Now, we focus on case (i), the T-models with $k=0$ which are compatible with the equation of state of a generic ideal gas. From now on, they will be called {\em ideal T-models}, and we study in detail for them the five steps required in analyzing the physical reality of the solutions.


\subsection{Metric line element of the ideal T-models} 
\label{subsec-metric-T-ideal}

Firstly, we achieve the first step of our procedure by completing the integration of the ideal sonic condition S$^{\rm G}$. As a consequence of the  the constraints (i) for $c_i$, $c_2 = c_1 + c_3 = \frac12$, we can consider a constant $\gamma$ such that 
\be  \label{ki-gamma}
c_1 = \frac{\gamma - 1}{2 \gamma} \, , \qquad c_2 = \frac{1}{2} \, , \qquad  c_3 = \frac{1}{2 \gamma}    \, .
\ee
Then, taking into account definitions (\ref{abcd}) and that $k=0$, equations (\ref{ideal-eq-b}) state:
\be \label{p-phi-ppunt}
p = \frac34 (\gamma -1) \frac{\dot{\varphi}^2}{\varphi^2} \, , \qquad \dot{p} \dot{\varphi} = -\frac{2 \gamma}{\gamma-1} \varphi p^2 \, .
\ee
This first order differential system for the functions $p(t)$ and $\varphi(t)$ can be easily integrated and we get:
\be  \label{phi-p}
\varphi  = \left[\frac32 \kappa \gamma (t-t_0)\right]^{\frac{4}{3 \gamma}}, \quad  p = 3 \kappa^2(\gamma-1) \varphi^{-\frac32 \gamma}  ,
\ee
where $\kappa$ is an arbitrary non-vanishing constant. Note that the constant $t_0$ determines an origin of time and can be taken as zero. Likewise, the change of the metric function $\varphi$ for a positive constant factor leaves the metric unchanged because it can be knocked out by changing the coordinates $x$ and $y$ for the square root of this factor. Nevertheless, the sign of the constant $\kappa$ determines the sign of the derivative of $\varphi$, $\dot{\varphi} = 2 \kappa \varphi^{1-\frac{3 \gamma}{4}}$. Consequently, $\kappa >0$ for expanding models, and then $t >0$. And for contracting models $\kappa < 0$, and then $t<0$. 

Now we are going to determine the metric function $\omega^2 = [\omega_1(t) + \omega_2(t) Q(r)]^2$. A straightforward calculation shows that, for $k=0$, $\omega_2 = \sqrt{\varphi}$ is a solution of equation (\ref{eq-T-2}). And $\omega_1 = \sqrt{\varphi} \alpha$ is a solution to this equation if, and only if, $\alpha = \alpha(t)$ fulfills $\dot{\alpha} = C \varphi^{-\frac32}$. Then, we can easily determine $\alpha(t)$ if we use the expression (\ref{phi-p}) for $\varphi(t)$, and then $\omega^2 = \varphi(t)[\alpha(t) + Q(r)]^2$. Note that, being $Q(r)$ an arbitrary function, the metric expression is invariant if we change $\alpha$ by an additive constant and a factor (changing appropriately the function $Q$ and the coordinate $r$). Finally, we arrive at:
\begin{proposition} \label{prop-T-ideal-metric}
The ideal T-models have a metric line element of the form:
\begin{equation} \label{metric-T-ideal}
ds^2= -dt^2 + \varphi(t)([\alpha(t) + Q(r)]^2 dr^2 + dx^2+dy^2),
\end{equation}
where $Q(r)$ is an arbitrary function and 
\begin{equation} \label{phi-alpha}
\varphi(t) = |t|^{\frac{4}{3 \gamma}}   ,  \qquad \alpha(t) = \cases{
 |t|^{1- \frac{2}{\gamma}}  , \, \qquad {\rm if} \ \ \, \gamma \not= 2  \cr 
  \ln |t|  , \quad \quad \ \ \,  {\rm if} \quad  \gamma = 2}
\end{equation}
The time coordinate takes values either in the interval $t>0$ (expanding models) or in the interval $t<0$ (contracting models).
\end{proposition}

On the other hand, from (\ref{phi-alpha}) and expression (\ref{expansion-T-2}), the expansion of the fluid flow takes the expression:
\begin{equation} 
\label{expansion-T-ideal}
\begin{array}{c}
\hspace{0cm}\displaystyle  \theta =\frac{2}{\gamma\,  t}  \Big(1 + \frac12 \delta \Big) , \quad \ \delta \! = \! \delta(t,r)\! \equiv \! \frac{\tilde{\alpha}(t)}{\alpha(t) \! + \! Q(r)}  , \\[4mm]  \quad \tilde{\alpha}(t) = \cases{\!(\gamma\!-\!2) \alpha(t),   \quad  \, {\rm if} \ \  \gamma \not= 2 \ \ \cr 
\!  2,   \qquad \qquad \quad   \,  {\rm if} \ \    \gamma = 2 \  \ }
\end{array}
\end{equation}
where $\alpha(t)$ is given in (\ref{phi-alpha}).


\subsection{Hydrodynamic quantities: energy density, pressure and speed of sound} 
\label{subsec-hydro-T-ideal}

Now we carry out the second step by obtaining the coordinate dependence of the hydrodynamic quantities $\rho$, $p$, and  the indicatrix function $c_s^2 = \chi(\pi)$. From the expressions (\ref{pressure-T-1}) and (\ref{density-T-1}) we can obtain the time dependence of the pressure and the energy density by taking $\omega_2 = \sqrt{\varphi}$ and $\omega_1 = \sqrt{\varphi}\alpha$ and making use of (\ref{phi-alpha}).  

On the other hand, the indicatrix function $\chi(\pi)$ can be determined from (\ref{chi-Tmodels}) by taking into account  (\ref{ABCci}) and (\ref{ki-gamma}). Then, we obtain the following:
\begin{proposition} \label{prop-T-ideal-hydro}
For the ideal T-models {\em (\ref{metric-T-ideal}-\ref{phi-alpha})} the pressure $p$ and the energy density $\rho$ take the expression:
\begin{eqnarray} \label{pressure-T-ideal}
p = \frac{4(\gamma-1)}{3 \gamma^2}\, \frac{1}{t^2}   \,  ,  \\[1mm]
\rho = \frac{4}{3 \gamma^2}\, \frac{1}{t^2}   [1 + \delta(t,r)]  \, ,
\label{density-T-ideal}
\end{eqnarray}
where $\delta(t,r)$ is given in {\em (\ref{expansion-T-ideal})}.
\ \\
And the square of the speed of sound is given by:
\be \label{chi-T-ideal}
c_s^2  = \chi(\pi)  \equiv \frac{2 \, \gamma \, \pi^2}{ (\pi + 1) (\pi + \gamma-1)} \, , \qquad \pi \equiv \frac{p}{\rho} \, .
\ee
\end{proposition}
\begin{table*}[t]
\begin{tabular}{cllll}
\hline \\[-4.5mm] \hline
  & \qquad \ \ $\alpha(t)$  & \  \qquad E$^{\rm G}$  &\qquad  \ $t$ &\qquad \quad \ $\delta$  
  \hspace{-10mm} \phantom{\LARGE $(\frac{A}{B})$} 
  \\[0mm]
 \hline
\hspace{-5mm}     &\ \ \quad   &  \ \qquad $Q < - \alpha $  &\qquad \  $]t_1,  \infty[ $  & \qquad \quad \ $> 0$, \ \quad  $\infty \searrow 0$ \phantom{\Large $(\frac{A}{B})$}
\\[-6mm] 
\hspace{-3mm} \quad   $\gamma < 2$  & \qquad \ \ $\infty \searrow 0$   &   &
\hspace{-20mm}  \phantom{\large $(\frac{A}{B})^{\Big(C \Big)}$}  
 \hspace{3cm}\\[-5mm]
 %
\hspace{-1mm}   &\    &  \ \qquad $Q > 0$  &\qquad \ $]0,  \infty[ $  & \qquad \quad \ $< 0$, \ \quad  $(\gamma\!-\!2) \nearrow 0 \ $    
\hspace{-20mm}  \phantom{\large $(\frac{A}{B})^{\Big(C \Big)}$}  
\\[1mm] \hline
\hspace{-3mm} \quad   $\gamma = 2$  & \qquad  \   $-  \infty \nearrow  \infty$   &\qquad $- \alpha < Q $  &\qquad  \  $]t_1,  \infty[ $  & \qquad \quad \ $> 0$, \ \quad  $\infty \searrow 0$
\hspace{-20mm}  \phantom{\large $(\frac{A}{B})^{\big(C \big)}$}  
 \hspace{3cm}\\[-4.3mm]
\\[2mm] \hline
\hspace{-3mm} \quad   $\gamma > 2$  &\qquad \ \  $ 0 \nearrow \infty $   & \qquad $- \alpha < Q < 0 $  &\qquad  \  $]t_1,  \infty[ $  &   \qquad \quad \ $> 0$, \ \quad  $\infty \searrow \gamma\!-\!2$
\hspace{-20mm}  \phantom{\large $(\frac{A}{B})^{\big(C \big)}$}  
\\[2mm] \hline 
\\[-3mm] \hline
\end{tabular}
\caption{This table provides, for the different values of the parameter $\gamma$: (i) the space-time region ${\cal R}_1$ where the energy conditions E$^{\rm G}$ hold (third column); (ii) the time interval where E$^{\rm G}$ is kept for a given $Q(r)$ (fourth column); (iii) the sign of the energy density contrast $\delta$, and the interval where it takes values (fifth column).}
\label{table-1}
\end{table*}


\subsection{Curvature singularities and space-time domains} 
\label{subsec-singularities}


Proposition \ref{prop-T-ideal-hydro} shows that the ideal T-models have a curvature singularity at $t=0$ and, when $Q(r) < 0$, another one at $\alpha(t)+Q(r)=0$. We briefly analyze them for the expanding models (for the contracting models the study is similar). 

The existence of these kind of singularities has been already remarked by several authors in the homogeneous case $\omega=\omega(t)$ (KCKS metrics). Kantowski \cite{Kantowski} pointed out that: (i) when $\varphi(t_0)=0$, the metric line element on the sphere (plane or hyperboloid) vanishes at $t=t_0$, and we have infinite energy density and pressure, and (ii) when $\omega(t_1)=0$, the one-dimensional metric line element $\omega^2 \dif r^2$ vanishes at $t=t_1$, and we have infinite energy density. On the other hand, Collins \cite{Collins} showed that, under the energy conditions (\ref{e-c}) and the first compressibility condition in (\ref{cc-1-chi}), the KCKS perfect fluid solutions are geodesically incomplete.

In the non-homogeneous case $\omega = \omega(t, r)$, we have also these curvature singularities, but the second one is not simultaneous for the co-moving observer. Now, the collapsing time depends on $r$, $t_1=t_1(r)$. 

In our ideal T-models we have $\omega = \sqrt{\varphi(t)}[\alpha(t)+Q(r)]$ and, consequently, $\omega=0$ when $\varphi=0$. Thus, the full line element of the 3-spaces $t= constant$ vanishes at $t=0$, and we have a big bang singularity. Both, energy density and pressure diverge at $t=0$. On the other hand, if $t_1=t_1(r)$ is such that $\alpha(t_1) + Q(r) =0$, the metric distance on the coordinate lines of the coordinate $r$ vanishes, and we have a singularity with a divergent energy density at $t = t_1$. 

This analysis shows that we have two disconnected space-time domains defined by:
\be
\begin{array}{l}
{\cal R}_0 = \{t >0 ,\quad  \alpha(t) + Q(r) < 0\} \, , \\[2mm]
{\cal R}_1 = \{t >0 ,\quad  \alpha(t) + Q(r) > 0\} \, .
\end{array}
\ee
Note that when $Q(r) > 0\ \ \forall r$, ${\cal R}_0= \emptyset$.


\subsection{Ideal T-models: analysis of the solutions and energy conditions} 
\label{subsec-energy-c}

The energy conditions E$^{\rm G}$ given in (\ref{e-c-G}) imply $p>0$. Then, the expression (\ref{pressure-T-ideal}) for the pressure means that, necessarily, $\gamma >1$. Note that we have a flat FLRW limit by taking $\alpha =0$ in the metric (\ref{metric-T-ideal}) (or, $\delta=0$ in the expressions of the expansion and energy density). In this limit we have a barotropic evolution of the form $p=(\gamma -1) \rho$. These FLRW models fulfill the energy condition (\ref{e-c-G}) when $\gamma < 2$, and they are the so-called $\gamma$-law models \cite{Assad-Lima}. The inhomogeneous models with $\gamma < 2$ belong to the the Szekeres-Szafron ideal singular models studied in \cite{CFS-PSS}. Nevertheless, in our inhomogeneous T-models with $\gamma \geq 2$ there may be regions where the energy conditions meet. We will also study them here.

Note that $\gamma$ is a thermodynamic parameter that defines the equation of state (\ref{chi-T-ideal}) and set the time dependence of the metric (see (\ref{metric-T-ideal}-\ref{phi-alpha})). The metric also depends on an arbitrary real function $Q(r)$ which determines the inhomogeneity. If $Q=constant$, then the metric is an (homogeneous) KCKS model. 

\begin{table*}[t]
\begin{tabular}{cllll}
\hline \\[-4.5mm] \hline
& $\qquad Q(\rho,p)$  &   $ \qquad \quad n(\rho, p) $  & \qquad \quad $\lambda_1(p)$  &\quad  \quad $\lambda_2(p) \quad $   
  \hspace{-10mm} \phantom{\LARGE $(\frac{A}{B})$} 
  \\[2mm]
  \hline
\hspace{2mm}  $ \gamma \not=2$  &   $  \qquad   \displaystyle \frac{\tilde{K}(\rho - p)}{\rho(\gamma-1)  - p}\,  p^{\frac{2-\gamma}{2 \gamma}}$ &  $ \quad \qquad \displaystyle \frac{\rho(\gamma-1)  - p}{K N(Q) \sqrt{p}}$  &  $ \quad \qquad l_1 \sqrt{p} $  &
\quad \quad $ l_2 \, p^{1 - \frac{1}{\gamma} \quad }$   
\hspace{-20mm}  \phantom{\Large $ \displaystyle [(\frac{A}{B})^{B}]$}  
\\[4mm] \hline
\hspace{2mm} $\gamma =2$  & $ \qquad  \displaystyle \frac12  \ln(3p) +  \frac{2p}{\rho-p} $ &
 $  \qquad \quad  \displaystyle \frac{\sqrt{3}\,(\rho - p)}{2 N(Q) \sqrt{p}}$  & \quad  \qquad $ \displaystyle  \sqrt{\frac{p}{3}} \, [2 - \ln(3p)] $ & \quad \quad $ \displaystyle  \frac{2}{\sqrt{3}} \sqrt{p} \quad $  
\hspace{-16.5mm}\phantom{\Large $ \displaystyle [(\frac{A}{B})^{B}]$}  
\\[4mm] \hline 
\\[-3mm] \hline
\end{tabular}
\caption{Thermodynamic schemes of the ideal T-models. This table offers the mass density $n(\rho,p)$ and the functions $Q(\rho,p)$ and $\lambda_i(p)$ that determine the specific entropy $s(\rho,p)=s(Q)$ and the temperature $\Theta(\rho,p) =  \ell(Q) \lambda_1(t) + m(Q) \lambda_2(t)$, with $\ell(Q)$ and $m(Q)$ given in (\ref{ell-m}). The constants $\tilde{K}$, $K$, $l_1$ and $l_2$ depend on the parameter $\gamma$ as $\tilde{K}\equiv  -(\gamma-1) \hat{\gamma}^{1- \frac{2}{\gamma}}$, $K\equiv(\gamma-2) \hat{\gamma}$, $l_1\equiv  2 \hat{\gamma}$ and $l_2 \equiv \frac{4}{3 \gamma} \hat{\gamma}^{\frac{1}{\gamma}-1}$, where $\hat{\gamma} \equiv \frac{2 \sqrt{\gamma-1}}{\sqrt{3} \gamma}$.} 
\label{table-2}
\end{table*}

If we denote the energy density of the FLRW limit as $\rho_F$, then we have $\rho = \rho_F(1 + \delta)$ and $p = (\gamma-1) \rho_F$. Thus, the function $\delta = \delta(t,r)$ given in (\ref{expansion-T-ideal}) is the energy density contrast with respect to the FLRW limit. Nevertheless, note that it is not the energy density contrast with respect to a homogeneous background (the KCKS limit acquired when $Q= constant$). 

With the notation introduced above, we have $\rho - p = \rho_F(2- \gamma + \delta)$. Consequently, the solution meets the energy conditions if, and only if, $\gamma > 1$ and $\delta > \gamma - 2$. The space-time regions where this last inequality holds strongly depend on whether $\gamma$ is greater than, equal to, or less than 2. The analysis of each case shows different behaviors summarized in Table \ref{table-1}. We only develop the expanding models ($t>0$) in detail. The behavior of the contracting models ($t<0$) can then be obtained from the expanding ones by exchanging the future for the past.

The energy conditions E$^{\rm G}$ involve the metric functions $\alpha(t)$ and $Q(r)$, and they only hold in the space-time domain ${\cal R}_1$. Moreover, the function $Q(r)$ can be chosen such that there is always a time $t_1$ where the E$^{\rm G}$ hold, and then they also hold  for later times. The only models with a negative $\delta$ can take place when $\gamma <2$ and $Q>0$. Then, the time coordinate covers its entire domain $t>0$ (that is, region ${\cal R}_1$, since ${\cal R}_0= \emptyset$), and $\delta$ increases from a finite negative value (at early times) to zero at later times (the model approaches the $\gamma$-law FLRW limit). Models with $\gamma < 2$ and a positive $\delta$ also approach the FLRW limit for later times. 
Models with $\gamma \geq 2$ have, necessarily, a positive $\delta$. When $\gamma=2$ the solution approaches the shift ($\rho=p$) FLRW model at later times. And, if $\gamma >2$, the energy density contrast decreases from large values and approaches a positive value for later times. 

Note that, for each model, the sign of the energy density contrast does not change throughout the space-time domain where the energy conditions E$^{\rm G}$ hold. Nevertheless, a suitable election of the function $Q(r)$ can model regions with an excess or a lack of energy density (with respect to a homogeneous KCKS background defined by a constant value of the function $Q$). 

It is worth remarking that the expansion (\ref{expansion-T-ideal}) of our inhomogeneous model has the same sign as the FLRW limit when $\delta > -2$. This occurs for the models with a positive energy density contrast, but also for $\delta <0$ in the domain where the energy conditions hold. This fact justifies that we speak of expanding models when $t>0$, and of contracting models when $t<0$.


\subsection{Ideal T-models: compressibility conditions {\rm H}$_1^{\rm G}$}
\label{subsec-compress-T-ideal}

The compressibility conditions H$_1^{\rm G}$ have been studied in \cite{CFS-PSS} for the Szekeres-Szafron ideal singular models. The indicatrix function in that case is of the form (\ref{chi-T-ideal}) with $1 < \gamma < 2$. Thus, we can now follow the same reasoning, which is also valid for $\gamma \geq 2$, and we obtain the same result for the ideal T-models as that obtained in \cite{CFS-PSS}. Namely, we have the following: {\it the ideal T-models  {\rm (\ref{metric-T-ideal}-\ref{phi-alpha})} fulfill the compressibility conditions ${\rm H}_1$  provided that they fulfill the energy conditions} E$^{\rm G}$, {\em that is, in the space-time domain} ${\cal R}_1$.

\begin{table*}[t]
\begin{tabular}{llllll}
\hline 
\\[-3mm] \hline
  & \qquad   $n(\rho, p) $  &\ \qquad  $\Theta(\rho, p)$  &\ \ \qquad  $s(\rho, p) $ & \qquad H$_2$  
  \hspace{-10mm} \phantom{\Large $(\frac{A}{B})$} 
  \\[1mm]
  \hline
 %
$\quad \gamma = 4/3$  & \qquad   $ \displaystyle \frac{(\rho - p)^{2}}{\rho - 3 p} $ & \  \qquad $ \displaystyle \frac{p}{ \tilde{k} \, n(\rho, p) } $  &\ \ \qquad  $  \displaystyle  s_0 \! + \! \tilde{k}  \ln \! \left[ \frac{1}{p}\!  \left[\frac{\rho - 3p}{\rho-p}\right]^{\!4} \right] $  & \qquad  $ \displaystyle \pi \in ]  \pi_1 , \frac13 [$  
\hspace{-18mm}  \phantom{\Large $ \displaystyle  (\frac{A}{B})^{(C)}$}  
\\[4mm] \hline
%
$ \quad \gamma = 2$ & \qquad $ \displaystyle  (\rho - p) \exp\left\{\frac{2 p}{\rho-p}\right\} $ &\  \qquad
 $ \displaystyle  \frac{p}{ \tilde{k} \, n(\rho, p) }$  &\ \  \qquad  $  \displaystyle  s_0 \! - \! \tilde{k} \left[ \ln  p + \frac{4 p}{\rho-p}\right] $  & \qquad  $\pi \in ]  \pi_2 , 1 [ $   
\hspace{-5.5mm} \phantom{\Large $ \displaystyle  (\frac{A}{b})$} 
\\[3.5mm] \hline 
\\[-3mm] \hline
\end{tabular}
\caption{This table provides, for the T-models with $\gamma =4/3$ and $\gamma=2$, the explicit expression of the matter density $n$, the temperature $\Theta$ and the specific entropy $s$ in terms of the hydrodynamic quantities $\rho$ and $p$ for the generic ideal gas thermodynamic scheme. Last column shows the  constraints imposed by the compressibility condition H$_2$: $\pi_1 = \frac{1}{13} (\sqrt{17}-2) \approx 0.16$, and $\pi_2 = \frac{1}{7} (2\sqrt{2}-1) \approx 0.26$.}
\label{table-3}
\end{table*}


\subsection{Thermodynamic schemes of the ideal T-models}
\label{subsec-scheme-T-ideal}

In the last two previous subsections we have acquired step 3 in analyzing the physical meaning of the ideal T-models. Now we can perform step 4 by particularizing the general study of the thermodynamic schemes presented in subsection \ref{subsec-scheme-Tmodels}. Note that the thermodynamic quantities depend on the metric functions $\varphi(t)$, $\omega_1(t)$ and $\omega_2(t)$, and on two functions, $N(Q)$ and $s(Q)$, of the metric function $Q(r)$. The former now take the expression $\omega_1 = \sqrt{\varphi} \alpha$, $\omega_2 = \sqrt{\varphi}$, with $\varphi(t)$ and $\alpha(t)$ given in (\ref{phi-alpha}). And each choice of the latter determines a specific thermodynamic scheme with a specific entropy and a mass density given in (\ref{s-n-Tmodels}), and a temperature given in (\ref{T-Tmodels}-\ref{ell-m}-\ref{lambda_i}). 

Thus, we can determine the thermodynamic quantities as a function of state depending on the hydrodynamic quantities $\rho$ and $p$ if we obtain the functions $Q(\rho,p)$ and $\lambda_i(p)$ given in (\ref{Q(prho)}) and (\ref{lambda_i}). Note that $Q(\rho,p)$ can be obtained from (\ref{expansion-T-ideal}-\ref{pressure-T-ideal}-\ref{density-T-ideal}). Table \ref{table-2} collects these expressions distinguishing the cases $\gamma\not=2$ and $\gamma=2$.

Then, we have that a particular ideal T-model admits a different thermodynamic interpretation for each choice of the functions $N(Q)$ and $s(Q)$. In \cite{CFS-PSS} we have studied in detail three thermodynamic schemes associated to the Szekeres-Szafron singular models that also apply for the ideal T-models when $\gamma < 2$: models with a generic ideal gas thermodynamic scheme, the Lima-Tiomno \cite{Lima-Tiomno} models, and the models with the temperature of the FLRW limit. In these three cases, step 5 of our approach has been analyzed: the positivity conditions P hold, and the compressibility condition H$_2$ holds in a wide space-time domain. All these results are summarized in Table 2 of \cite{CFS-PSS}.
A detailed study of different thermodynamic schemes for any $\gamma$ falls outside the scope of this paper. Here, we will limit ourselves to outlining some qualities of the scheme that allows us to interpret the solutions $\gamma = 4/3$ and $\gamma=2$ as generic ideal gases in local thermal equilibrium. In \cite{CFS-LTE} we presented an algorithm that provides all the thermodynamic quantities of the ideal gas scheme when the indicatrix function $\chi = \chi(\pi)$ is known. 

If we consider the expression (\ref{chi-T-ideal}) for $\chi(\pi)$ when $\gamma = 4/3$ or $\gamma=2$ and we apply this algorithm we obtain the thermodynamic schemes summarized in Table \ref{table-3}. On the other hand, we know that, for the ideal gas schemes, the compressibility condition H$_2$ holds if, and only if, the indicatrix function $\chi(\pi)$ fulfills \cite{CFS-CC}:
\be
\hspace{-5mm} {\rm H}^{\rm G}_2 : \qquad \qquad    \xi \equiv (2 \pi + 1) \chi(\pi) - \pi > 0 \, .
\ee
For the $\chi(\pi)$ of the ideal gas T-models this inequality holds in an interval $]\pi_m, 1[$ as the last column in Table \ref{table-3} shows. It is worth remarking that when $\gamma =4/3$ (and similarly, for any $\gamma <2$) the thermodynamic variables are not defined for $\pi=1/3$ (similarly, for $\pi = \gamma-1$). The two resulting subintervals are related with the two different cases where the energy conditions hold when $\gamma<2$ (see Table \ref{table-1}). Indeed, when $Q < - \alpha$ we have $\delta >0$ and then $\pi < \gamma-1$; and when $Q >0$ we have $\delta <0$ and then $\pi > \gamma-1$. In Table \ref{table-3}, for the models $\gamma =4/3$, we have only considered the expression of the thermodynamic variables when  energy density contrast is positive. In this case, we have $\pi <1/3$, and in the limit $\chi(1/3) = 1/3$, $\chi'(1/3) = 1/2$, the same values that the Synge gas \cite{Synge}. Thus, this ideal gas scheme appears to be a good approximation to a relativistic gas.

Note that the ideal gas thermodynamic schemes considered in Table \ref{table-3} can also be obtained from the generic thermodynamic schemes in Table \ref{table-2} by considering a particular choice of the functions $s(Q)$ and $N(Q)$: for $\gamma=4/3$, $s(Q) = s_0-\frac14 \ln|Q|$, $N(Q)= -\frac23 Q^{-2}$; for $\gamma=2$, $s(Q) = s_0-\ln 3+2 Q $, $N(Q)= \frac23\exp\{-Q\}$.


\section{The McVittie-Wiltshire-Herlt solution and its generalizations}
\label{sec-McVittie}

As commented in section \ref{sec-Tmodels}, Herlt \cite{Herlt} proposed a method to get an inhomogeneous T-model from a known homogeneous KCKS metric, and he then applies this method to generalize the (spherically symmetric) McVittie and Wiltshire solution \cite{McVittie}. In the canonical form (\ref{metric-T-1}) they take the time coordinate $\tau$ such that $Y=\sqrt{\varphi}= \tau$, and they look for the model with $\omega =  \tau^n \equiv \omega_1(\tau)$. Then, the field equation (\ref{eq-T-1}) determines the function $v(\tau)$: 
\be
v(\tau) = \frac{1}{n^2 -1} + C_0 \, \tau^{-2(n+1)} , 
\ee
which gives the McVittie-Wiltshire solution. Then, substituting this expression in (\ref{eq-T-1}) we obtain an equation for $\omega(t)$. We know the particular solution $\omega_1(t)$, and then we can formally determine another one, $\omega_2(t)$, in terms of an integral \cite{Herlt}. When $C_0 =0$ this integral can be explicitly calculated and one obtains $\omega_2(\tau) = \tau^{-n}$. We name this specific solution the McVittie-Wiltshire-Herlt T-model (see also \cite{kramer}).

We can easily recover the McVittie-Wiltshire-Herlt (MWH) solution by working with the proper time $t$ of the Lagrangian observer. When $C_0=0$ we have that $\dif t = v^{-1/2} \dif \tau = \sqrt{n^2-1} \dif \tau$ and, consequently, $\varphi = Y^2 = f_0 \, t^2$, with $f_0^{-1} = n^2 - 1 >0$.

In this section we study the T-models (\ref{metric-T-2}-\ref{eq-T-2}) with $\varphi = f_0 \, t^2$. In this way, we generalize the MWH T-model to any curvature $k=0, \pm 1$. Moreover, we analyze for these solutions the macroscopic necessary condition for physical reality. 


\subsection{Metric line element}

If we take $\varphi(t) = f_0 \, t^2$, then the field equation (\ref{eq-T-2}) becomes:
\be
t^2\,  \ddot{\omega} + t \, \dot{\omega} - k_0 \, \omega = 0  \, , \qquad k_0 \equiv 1+ \frac{k}{f_0} \, .
\ee
A straightforward calculation shows that this equation admits a solution if $k_0 >0$, and two independent ones are:
\be \label{omegues}
\omega_1(t) = t^n , \quad \omega_2(t) = t^{-n}  , \qquad  n = \sqrt{k_0} > 0 \, .
\ee
Then, the metric line element of the generalized MWH T-models takes the form (\ref{metric-T-2}), where $\omega_i(t)$ are given in (\ref{omegues}) and:
\be \label{varphi}
\varphi(t) = f_0 \, t^2 > 0 \, .
\ee
Note that, if $k=1$, then $n>1$, and we recover the MWH solution, if $k=-1$, then $n<1$, and if $k=0$, then $n=1$.

It is worth remarking that the choice $\varphi(t) = f_0 \, t^2$ is equivalent to $\dot{\varphi}^2 = 4 f_0 \, \varphi$. Consequently, the case (ii) named in subsection \ref{subsec-ideal-equations} corresponds with the generalized MWH metrics.


\subsection{Pressure and energy density. Study of the energy conditions}

Now, we particularize the expressions of the pressure (\ref{pressure-T-2}) and the energy density (\ref{density-T-2}) for the generalized MWH T-models and we obtain:
\begin{eqnarray} \label{pressure-McV}
p = -  \frac{n^2}{t^2}   \,  ,  \\[1mm]
\rho = \frac{n^2}{t^2} \left[1 +   \frac2n \, \frac{t^{2n} - Q(r)}{t^{2n} + Q(r)} \right]  \, .
\label{density-McV}
\end{eqnarray}
Note that the pressure is negative and this fact disqualifies these solutions as ideal gas models. Nevertheless, it is known that continuous media with negative pressures exist and it is suitable to analyze the energy conditions for these models.

From the expressions (\ref{pressure-McV}) and (\ref{density-McV}) it follows that the first inequality of the energy conditions E given in (\ref{e-c}), $-\rho<p$, is equivalent to $(t^{2n} - Q)(t^{2n}+Q)^{-1}>0$. Both factors of this expression cannot be simultaneously negative, and both are positive if, and only if,
\be \label{e-cMcV}
|Q(r)| < t^{2n} \, .
\ee
On the other hand, the second energy inequality, $p\leq\rho$, holds if $\rho>0$, that is, if $(t^{2n} - Q)(t^{2n}+Q)^{-1}>- n/2$, which is a consequence of the above condition (\ref{e-cMcV}). Moreover, in this case the expression (\ref{expansion-T-2}) of the expansion becomes:
\be
\theta = \frac{2}{t} \left[1 +   \frac{n}{2} \, \frac{t^{2n} - Q(r)}{t^{2n} + Q(r)} \right]  \, .
\ee
Note that if the energy conditions hold, then the sign of the expansion depends on the sign of the time coordinate.

Thus, it follows that the generalized MWH solution fulfills the energy conditions in the spacetime domain defined by (\ref{e-cMcV}). Moreover, for a given time $t_1$, we can always choose the inhomogeneity function $Q(r)$ such that (\ref{e-cMcV}) holds. Then, for expanding models the energy conditions hold for $t>t_1>0$ (in the future); and, for contracting models ($t<0$) the energy conditions hold for $t<t_1<0$ (in the past).


\subsection{Speed of sound. Compressibility conditions}

We can obtain the indicatrix function $\chi(\rho,p)$, which gives the square of the speed of sound, by specifying the general expression (\ref{chi-Tmodels}) of the T-models for this case. Indeed, from the expressions (\ref{omegues}-\ref{varphi}) of the metric functions, and taking into account (\ref{abcd}-\ref{p-density-expansion}-\ref{rho-punt}), we can determine the functions (\ref{ABCcal}). Then, by substituting in (\ref{chi-Tmodels}) we obtain:
\be \label{chi-McV}
c_s^2  = \chi(\pi)  \equiv \frac{4 \, \pi^2}{ (\pi + 1) [(4-n^2) \pi - n^2]} \, , \quad \pi \equiv \frac{p}{\rho} \, .
\ee
Note that the indicatrix function is of the ideal gas type, in accordance with case (ii) of subsection \ref{subsec-ideal-equations}. Then, we can analyze the compressibility conditions H$^{\rm G}_1$ in the regions where the energy conditions meet. In this case we have $-1 < \pi <0$, and then the first inequality in (\ref{cc-ideal}), $0 <\chi$, implies $n^2 <  (4-n^2) \pi = (n^2-4)(-\pi) < n^2 -4$. This contradiction shows that compressibility conditions are not satisfied anywhere.

In summary, the McVittie-Wiltshire-Herlt solution is not a good model to represent a perfect fluid in local thermal equilibrium.


\section{Discussion}
\label{sec-discussion}

\subsection{Analysis of the results}

In this paper we have shown that, for the T-models, the field equations can be written as a differential equation that is linear for an adequate choice of the three unknown metric functions. One of these functions can be arbitrarily fixed with a specific choice of the time coordinate, and then the space of solutions is controlled by two real functions $\{\varphi(t), Q(r)\}$, which fully determine the gravitational field. Then, the hydrodynamic quantities of the fluid, unit velocity $u$, energy density $\rho$ and pressure $p$, are also fixed by the functions $\{\varphi(t), Q(r)\}$.

Each of these solutions can be furnished with a set of thermodynamic quantities, matter density $n$, entropy $s$, temperature $\Theta$ and specific internal energy $\epsilon$, constrained by the common thermodynamic laws. The richness of such thermodynamic schemes also depends on two arbitrary real functions $\{N(Q), s(Q)\}$, and they offer different thermodynamic interpretations of a given gravitational field $\{\varphi(t), Q(r)\}$. Here we have given the expression of the thermodynamic quantities $\{n, s, \Theta, \epsilon\}$ in terms of the four functions $\{\varphi(t), Q(r), N(Q), s(Q)\}$.

The thermodynamic study commented above is formal but it points out the solutions and the thermodynamics that are candidates to model a physically realistic perfect fluid in local thermal equilibrium. Complementary macroscopic physical requirement (energy and compressibility conditions and positivity of some thermodynamic quantities)  must be imposed on the thermodynamic solutions in order to obtain physically realistic models. Here we have imposed the compatibility of the thermodynamic solutions with the generic ideal gas equation of state, a condition that is only compatible with plane symmetry. Then, the metric function $\varphi(t)$ depends on a constant parameter $\gamma$, which fixes the function of state that gives the square of the speed of sound in terms of the energy density and the pressure. Moreover, we obtain for any $\gamma$ the space-time domains where the macroscopic physical constraints hold. 

We have also analyzed from a thermodynamic perspective the previously known MWH solution and we have shown its unsatisfactory physical meaning as a perfect fluid in local thermal equilibrium.


\subsection{Why there are no solutions that model a classical ideal gas?}
\label{sec-CIG}

Classical ideal gases are the ideal gases that also fulfill the classical dependence of the specific internal energy on the temperature, $\epsilon = c_v \Theta$. For them, the indicatrix function takes the form $\chi= \frac{\gamma \pi} {1 + \pi}$, where $\gamma \equiv 1 + \frac{\tilde{k}}{c_v}$ is the adiabatic index \cite{CFS-CIG}. The study undertaken in section \ref{sec-chi-pi} on the T-models compatible with the equation of state (\ref{gas-ideal}) of a generic ideal gas leads to an indicatrix function of the form (\ref{chi-Tmodels}). Thus, no solutions that model a classical ideal gas exist. 

It is worth remarking that we find a similar negative result in analyzing the ideal gas models belonging to the family of the Szekeres-Szafron solutions of class II, in both singular \cite{CFS-PSS} and regular \cite{CFS-RSS} models. In \cite{CFS-CIG} we have also searched for classical ideal gas solutions in the family of the R-models in geodesic motion, and the result has also been negative: the only solutions are the homogeneous ones (classical ideal gas FLRW models \cite{CFS-CIG}).

A question naturally arises: are these negative results a consequence of a more general basic result? The answer is affirmative. Indeed, in \cite{FS-KCIG} we have characterized the unit velocities of the classical ideal gas solutions of the hydrodynamic equations, and we have shown the following result: a geodesic and expanding time-like unit vector $u$ is the unit velocity of a classical ideal gas if, and only if, $u$ is vorticity-free and its expansion is homogeneous, that is, $u= - \dif t$ and $\theta=\theta(t)$.

Note that the Szekeres-Szafron solutions have a geodesic and expanding fluid flow. Consequently, only those with homogeneous expansion can be a candidate to model a classical ideal gas. But, for these metrics, homogeneous expansion is tantamount to barotropic evolution. Thus \cite{Krasinski, FS-SS}, for class II (and consequently in the limit admitting a G$_3$, the T-models) the metric is either a FLRW model or a KCKS solution; and, for class I (and consequently in the limit admitting a G$_3$, the geodesic R-models), the metric is necessarily a FLRW model.  

The generalized Friedmann equation for the classical ideals gas FLRW models has been presented in \cite{CFS-CIG}. These models have a specific barotropic equation of state $p = p(\rho)$ that follows by imposing an isentropic evolution. The study of the KCKS solutions that model similar physical properties is an ongoing work that will be presented elsewhere.

Note that the constraints on the kinematics of a classical ideal gas studied in \cite{CFS-CIG} are a consequence of the sole hydrodynamic equations and they do not depend on the field equations. This means that there are also no test solutions modeling a classical ideal gas that is comoving with the prefect fluid flow of the  non-homogeneous solutions quoted above.


\subsection{Work in progress}

The study of the thermodynamic T-models presented here further our understanding  of the physical meaning of this solutions but it also suggests new open questions that should be answered. The first one poses the possible thermodynamic interpretation of the homogenous limit of the T-models, the KCKS metrics. The results in subsection \ref{subsec-scheme-Tmodels} show that an isentropic evolution of each thermodynamic T-model leads to $Q(r)= constant$, that is, to a KCKS solution defined by a specific barotropic relation $p=p(\rho)$. Moreover, as commented in the subsection above, we will study the KCKS models that represent the isentropic evolution of a classical ideal gas, but these solutions are not the homogeneous limit of classical ideal gas inhomogeneous T-models.

On the other hand, very few explicit solutions of the T-model equation (\ref{eq-T-1}) are known. A deeper analysis of this equation is underway to find new physically reasonable solutions, and particularly, spherically symmetric ones.

Further research will address a similar analysis for the R-models. A general study of their thermodynamic interpretation has yet to be done. Only partial results are known at present. The classical ideal gases in geodesic motion have been considered in \cite{CFS-CIG}, and the ideal gas Stephani universes were examined in \cite{CF-Stephani}.


\begin{acknowledgements}
This work has been supported by the Spanish Ministerio de Ciencia, Innovaci\'on y Universidades and the Fondo Europeo de Desarrollo Regional, Projects PID2019-109753GB-C21 and PID2019-109753GB-C22, the Generalitat Valenciana Project AICO/2020/125 and the University of Valencia Special Action Project  UV-INVAE19-1197312. 
\end{acknowledgements}


\bibliography{Termo-T_PRD}

\end{document}